\documentclass[a4paper,10pt,twocolumn]{article}

\usepackage[utf8]{inputenc}
\usepackage{epsfig}
\usepackage{graphicx}
\usepackage{subfigure}
\usepackage{multirow}
\usepackage{dcolumn}
\usepackage{bm}
\usepackage{amssymb}
\usepackage{pifont}
\usepackage{amsmath}
\usepackage{times}
\usepackage[font=small,labelfont=bf,textfont=sl]{caption}
\usepackage{enumitem} 
\usepackage[dvips]{color}
\usepackage{abstract}
\usepackage[compact]{titlesec}
\usepackage{float}
\usepackage[numbers,sort&compress]{natbib}  
\usepackage{authblk}                        

\titleformat*{\section}{\bfseries\raggedright\small} 
\titleformat*{\subsection}{\itshape\raggedright\small} 
\titlespacing{\section}{0pt}{20pt}{5pt} 
\titlespacing{\subsection}{0pt}{14pt}{5pt} 

\definecolor{darkred}{rgb}{0.75,0.1,0.1}
\definecolor{darkgreen}{rgb}{0.3,0.5,0.1}

\newcommand{\HRule}{\rule{\linewidth}{0.1mm}}

\date{}

 \title{Impact of wettability correlations on multiphase flow through porous media}

\author[1,*]{\rm M. S. de La Lama}
\author[1]{\rm M. Brinkmann}
\affil[1]{\small \it Department of Dynamics of Complex Fluids, Max-Planck Institute for Dynamics and Self-Organization \\

                 Am Fassberg 17 D-37073, G\"ottingen, Germany}

\begin{document}

\twocolumn[
\maketitle

\begin{onecolabstract}
  In the last decades, significant progress has been made in
  understanding the multiphase displacement through porous media with
  homogeneous wettability and its relation to the pore geometry. However, the
  role of wettability at the scale of the pore remains still
  little understood. In the present study the displacement of
  immiscible fluids through a two-dimensional porous medium is
  simulated by means of a mesoscopic particle approach. The substrate
  is described as an assembly of non-overlapping circular disks whose
  preferential wettability is distributed according to prescribed
  spatial correlations, from pore scale up to domains at system
  size. We analyze how this well-defined heterogeneous wettability
  affects the flow and try to establish a relationship among
  wettability-correlations and large-scale properties of the
  multiphase flow. 
  
   \vspace*{10pt}

   \HRule \vspace*{10pt}
\end{onecolabstract}
]

\footnotetext[1]{E-mail for correspondence: m.s.delalama@ds.mpg.de}

\section{Introduction}

The displacement of immiscible fluids through porous media is subject
of scientific interest, as it is involved in many industrial and
technological applications, such as oil recovery, water flows, and soil
treatment. Understanding the dynamics of this process at the pore scale is
crucial but, on the other hand, very challenging due to the many
involved parameters, from the competition between viscous and
capillary forces to the wetting properties and porous structure of the
substrate.

In the last decades visualization experiments using simplified
micromodels of porous media have provided a detailed description of
immiscible flow~\cite{lenormand_1988,frette_1997,theodoropoulou_2005,berejnov_2007,cottin_2011}.
A number of dynamical regimes are distinguished in terms of
capillary number (ratio of viscous to capillary forces), viscosity
ratio (ratio of the viscosity of the injected fluid to the viscosity
of displaced fluid), and wetting conditions.

In the limit of high injection rates, the front advance critically
depends on the viscosity ratio of the fluids.
When the injected fluid has a lower viscosity than the displaced
fluid, the situation is highly unstable and ramified viscous fingers
are observed.
In the opposite case, when the injected liquid has a higher viscosity,
the front exhibits a stable displacement with a steady width.
However, in the limit of very slow injection rates the displacement
depends entirely on the capillary forces.
In this limit, the pressure drop due to the viscous flow field can be
neglected and the front roughens due to capillary pressure
fluctuations at the pore level.
This \textit{capillary fingering} regime has been successfully
described by means of statistical models as
invasion-percolation~\cite{wilkinson_1983,lenormand_1990,cottin_2010} and some more
sophisticated pore-network models~\cite{blunt_2004,ferer_2007,tora_hansen_2012}.

On the other hand, the development of analytical and computational
techniques for fluid dynamic simulations have provided a great advance
in the numerical approach to this phenomenon
~\cite{hilfer_1998,bogoyavlensky_2001,babchin_2008,riaz_2007}. However,
the continuous description is not sufficient to understand the multiphase
flow at the pore scale due to the wide range of scales involved in this
process.  To do so, one has to comprehend the interfacial dynamics -
especially the fluctuations generated at the contact line by the
substrate heterogeneities- and, from this microscopical information,
build up a coherent macroscopic description of the flow.

Mesoscopic simulation techniques provide a robust numerical approach
with high enough computational efficiency at the scale of interest.
The fundamental idea is to design simplified kinetic models that
coarse-grain the essential microscopical physics into a mesoscopic
description, so that the averaged properties obey the
macroscopic transport equations, in particular Navier-Stokes and mass
diffusion equations. These mesoscale algorithms properly reproduce the
complex behavior of systems like colloidal suspensions or
microemulsions, and have the major advantage of dealing with changing
interfacial topologies in a natural way.
The most extended ones are direct Monte-Carlo simulations (MC)~\cite{bird_book},
Lattice Boltzmann Method (LBM)~\cite{succi_book}, Dissipative-Particle
Dynamics (DPD)~\cite{hoogerbrugge_1992} and Multi-Particle Collision models (MPC)~\cite{malevantes_1999,gompper_2009}.

In the present paper we employ the MPC method, originally proposed by
Malevantes and Kapral~\cite{malevantes_1999}, to simulate the
displacement of immiscible fluids in a two-dimensional porous medium at
low capillary numbers. This particle mesoscopic approach is an
alternative simulation technique that, compared with the most extended
LBM, directly includes thermal fluctuations, allows to treat systems
with an arbitrary number of phases, and provides detailed information
about the dynamics at the fluid boundaries on any scale, as the
definition of the physical system is independent on the coarsening
process of the mesoscopic approach.

To deal with multiple phases we adopt the MPC multicolor
generalization proposed by Inoue \textit{et al.}~\cite{inoue_2004}.
We incorporate to the model the definition of relative adhesion of the
fluid phases to the solid phase, in order to control the wetting
conditions.  Our aim is to provide a numerical tool to study the
influence of wettability on immiscible multiphase flow at the pore
scale.  Although significant progress has been made toward
understanding the effect of homogeneous substrate wettability, the
role of heterogeneous wettability remains still little understood due
to the difficulty of controlling its spatial definition at the scale
of the pores~\cite{singhal_1976,salin_1990,hazlett_1998,bertin_1998}. Here, we
prepare different spatial correlations of local wettability in our
numerical simulations and analyze how they influence the dynamics,
trying to establish a relationship with large-scale properties of
the multiphase flow.

In the next Section \ref{sec:MPC_model}, we provide a description of
the MPC method including the wettability implementation. In Section
\ref{sec:system_parameters_and_thermal_fluctuations} we analyze the
relevant parameters interfacial tension and viscosity, and their
consistency with thermal fluctuations in terms of capillary waves. In
Section \ref{sec:flow_in_porous_media} we present our results on
multiphase flow through porous media analyzing the role of wettability
for both homogeneous and heterogeneous cases.

\section{MPC model}
\label{sec:MPC_model}

The fluid is modeled by a large number $N$ of point-like particles,
each with the same mass $m$ which move with a continuous distribution
of velocities. The dynamics of the particles consists in a sequence of
streaming and collision steps. In the streaming step, the particles
move deterministically during a time interval $\Delta t$
according to their individual velocity $v_i(t)$.

In order to introduce an interaction among particles, i.e. an exchange
of linear momentum, they are sorted into collision cells $\xi$. As we are
working in two spatial dimensions, we use a regular square grid of
typical size $a$ where the number of particles per cell, $N_{\xi}$,
fluctuates around an average value $N_c$. In every collision step, the
velocities of the particles are decomposed into the center of mass
velocity $\vec{u}_\xi$ of the particles in a cell and a remaining, fluctuational
part. An effective exchange of linear momentum between the particles
in cell is achieved through a rotation of the fluctuational velocity
components in each cell.  The velocity after the effective collision
step at time $t+\Delta t$ is:
\begin{equation}
  \vec{v}_i(t +\Delta t) 
  = \vec{u}_{\xi}(t)+\Omega [\vec{v}_i(t)-\vec{u}_{\xi}(t)].
  \label{eq:collision_step}
\end{equation}
As we assume molecular chaos, the rotation operator $\Omega$ in
eqn.~(\ref{eq:collision_step}) must be statistically independent in
each cell and in each time-step. Galilean invariance is guaranteed by
shifting the coarse-graining grid randomly before each collision
step~\cite{ihle_2001}. As linear momentum and mass is conserved, the
spatially averaged particle motion displays hydrodynamic behavior on
large length scales.  Furthermore it has been shown that there is an
\textit{H-}theorem for this algorithm~\cite{malevantes_1999}.

To deal with multiple immiscible fluid phases we adopt a variant of
the MPC algorithm proposed by Inoue \textit{et al.}~\cite{inoue_2004},
that induces phase segregation. This method has been already successfully
applied to study the distribution of droplets in bifurcating micro-channels~\cite{inoue_2006}. 
In this algorithm, each phase is assigned
a label (color) $c=1,...,N_{ph}$. At each time step and cell $\xi$ the color flux
\begin{equation}
  \vec{q}_c(\xi)
  =\sum_{i=1}^{N_{\xi}}(\vec{v}_i-\vec{u}_{\xi})\cdot \delta_{c,c_i} \ ,
\end{equation}  
and the color-gradient field 
\begin{equation}
  \vec{F}_c(\xi)=\sum_{j=1}^{N_{ph}} \kappa_{cj}\cdot \vec{\nabla}\,n_j
\end{equation}
are calculated, where $\vec {\nabla}\,n_j$ is the local gradient of the number density
$n$ of particles with color $j$ which is estimated from the number of
particles of the same color in the next-nearest neighbor cells. The
matrix $\kappa_{ij}$ is the \textit{interaction matrix} that defines
the cohesion/adhesion between fluid phases of color $i$ and $j$.
Then the multicolor potential energy is defined as
\begin{equation}
  U(\xi)=\sum_{c=1}^{N_{ph}} U_c(\xi)=\sum_{c} -\vec{q}_c(\xi)\cdot \vec{F}_c(\xi)
\end{equation} 
and the operator $\Omega$ is chosen such to minimize $U(\xi)$ (in the
case of 2-dimensions this simply means $\Omega(\theta_{\xi})$ with
$\partial U/ \partial \theta_{\xi}=0$, where $\theta_{\xi}$ is the
angle of rotation respect the center of the cell). Let us mention here
that this segregation algorithm generates a depletion layer at the
fluid-fluid interface that effectively reduces the transport of shear
stress through the fluid interface and, in particular, a finite slip
between the two fluid phases. However, as we are mainly working in the
regime of small capillary numbers, where normal stresses dominate the
dynamics at the fluid interfaces, this effect is not relevant. For a
more detailed description of the model, the reader is referred
to~\cite{inoue_2004}.

Thermostatting is required in any non-equilibrium MPC simulation due
to viscous heating. Here we apply a simple local profile-unbiased
thermostat that ensures control of the thermal fluctuations in the
bulk as well as on the solid walls while keeping unaffected the
velocity field~\cite{evans_1986}.  The relative velocities to the
center-of-mass velocity of each cell $\xi$ are rescaled after each collision step as
$\vec{v_i}=\lambda(\xi)\cdot\vec{v_i}$, where $\lambda(\xi)=k_BT N_\xi
d/\sum_{i=1}^{N_\xi}m\cdot (\vec{v_i}-\vec{u_\xi})^2$ with $d$
dimensions.

\subsection{Wettability implementation}
\label{subsec:wettability_implementation}

The multicolor model was originally proposed without specifying any
wetting condition~\cite{inoue_2004}. However, wettability plays a
pivotal role in immiscible flow through porous media and, hence, needs
to be thoroughly implemented in the algorithm. To this end we
developed a method that accounts for relative adhesion of two fluids
to the solid surface while, at the same time fulfilling the proper
fluid dynamic boundary conditions.

First, let us review the treatment of the fluid dynamic boundary
conditions for MPC algorithms.  In order to guaranty non-slip boundary
condition for the average fluid velocities, we employ the
generalization of the bounce-back rule for partially filled
cells~\cite{lamura_2001}. In the na\"ive formulation of the bounce back
rule, particles travel back into the direction of their incidence
after having collided with the solid boundaries. However, because the position
of the solid boundaries relative to the coarse-graining grid changes
between every interparticle collision step, it is necessary to add a
virtual phase resting in the walls. This virtual phase takes part in the
collision procedure to match the bulk particle density in the underfilled cells.
These wall-particles are generated before and removed after every interparticle
collision step, and guarantee a no-slip boundary condition at the
walls~\cite{lamura_2001}.

To define the wetting conditions we include the virtual wall-phase in
the definition of the interaction matrix, adding an extra term that
determines the wall interaction for each of the fluid phases present in the
system. By tuning the relative strength of each of them, we can
achieve different contact angles of the fluid interface at the solid
walls. In the following we will refer to a system of two phases,
namely fluid~(1) and fluid~(2), in contact with the wall~(3), and our
matrix is $\kappa_{ij}, i,j=1,2,3$.

In order to reduce the slip generated by the phase segregation
procedure, we set $\kappa_{13}=1$ that fixes the surface tension
$\sigma_{13}=0$. As there is no distinction between phase~(1) and (3)
there will be no slip for fluid phase~(1) at the wall (3).
Then, our interaction matrix will have the form
\begin{equation}
  \kappa = 
  \left( 
    \begin{array}{rrr}
      1 & -1   &  1 \\
      -1 &  1  & \kappa_{23} \\
      1 & \kappa_{23} & 1
    \end{array}  
  \right)
  \label{eq:interaction_matrix}
\end{equation}

According to the definition of the Young's contact angle, referred to
a droplet of fluid (2) resting on the wall and immersed in the ambient
fluid (1), we have $\sigma_{12}\cos\theta+\sigma_{23}=\sigma_{13}$
that is simplified to $\sigma_{12}\cos\theta+\sigma_{23}=0$ with the
condition $\kappa_{13}=1$. In the remainder of
this article we will focus on three particular cases: i) For
$\kappa_{23}=-1$ we find $\sigma_{23}=\sigma_{12}$ and therefore a
contact angle of $\theta=180^{\circ}$ while ii) $\kappa_{23}=1$ leads
to $\sigma_{23}=0$ and a contact angle of $\theta=90^{\circ}$.  iii)
In the limit $\kappa_{23}\gg 1$ we have $\sigma_{23}<0$ and therefore
$\theta\rightarrow 0^{\circ}$.

\begin{figure}[bht!]
  \begin{center}
    \epsfig{file=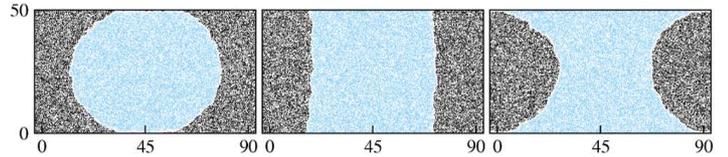,width=0.5\textwidth}
    \caption{Equilibrium contact angle for the three considered
      wetting conditions in a planar geometry (from left to right):
      \textit{non-wetting, neutral wettability} and
      \textit{total wetting}.}
    \label{fig:equilibrium_contact_angle}
  \end{center}
\end{figure}

In the following, as we denote the invading phase by (2), we will
refer to these cases as non-wetting ($\kappa_{23}=-1$), neutral
wettability ($\kappa_{23}=1$) and total wetting, for which we consider
$\kappa_{23}=2$. The introduction of a negative surface tension for
the wetting case allows us to avoid the formation of a depletion layer
originated by the phase segregation mechanism and avoid any slippage
of the invading phase (2) at the wall.  In
Fig.~\ref{fig:equilibrium_contact_angle} we show the equilibrium
configurations of a droplet of phase (2) between two solid parallel
walls for the three wetting conditions we consider.

\section{System parameters and thermal fluctuations}
\label{sec:system_parameters_and_thermal_fluctuations}

In this section we analyze how macroscopic quantities, like the
dynamic shear viscosity and interfacial tension, emerge from the
definition of our microscopic system parameters, that are the cell
size $a$, time step $\Delta t$, particle mass $m$, mean number density
$N_c$ defined as the average number of particles per cell, temperature
$temp$ (in $k_BT$ units), and the entries $\kappa_{ij}$ of the
interaction matrix. We normalize $\Delta t=1$, $a=1$, $m=1$, and set
$N_c=10$. For all the simulations we fix $temp=10^{-3}$, which
implies a mean free path for a particle in between two consecutive
collisions $\lambda=\Delta t \sqrt{temp/m}\approxeq 0.03$.

The behavior of a single phase does not depend on the strength of
the self-interactions among particles of the same species, so we
can normalize the diagonal terms of the interaction matrix to
$\kappa_{ii}=1$ without loss of generality.  Fitting
the velocity profile of a simple fluid in a planar channel to the
Poiseuille flow we estimate the value of the dynamic viscosity of the
fluid phase to be $\mu=0.750\,m/a \Delta t$ for the set of parameters
considered.

Since the interfacial tension the fluid phases (1) and (2) is not a
direct input for our algorithm but is controlled by the interaction
term $\kappa_{12}$, we estimate it by terms of the \textit{Laplace
  law}.  Initially a square droplet of fluid (1) is placed in the
center of a simulation box filled with fluid (2). After a transient
time, it relaxes to a configuration fluctuating around a
circular shape. We measure the radius $R$ of the averaged droplet
contour and the corresponding difference $\Delta P= P_{1}-P_{2}$
between the pressure in the droplet (1) and the in the ambient fluid
(2) for different values of the interaction term $\kappa_{12}$. The
pressure is computed by means of the microscopic definition of the
local stress tensor
\begin{equation}
  T_{\alpha\beta}
  =-\frac{1}{V}\sum_{i=1}^{N_c} m_i\cdot v_{i\alpha} \cdot v_{i\beta}
  -\frac{1}{V \Delta t}\sum_{i=1}^{N_c} 
  \Delta p_{i\alpha} \cdot r_{i\beta}~,
  \label{eq:stress_tensor}
\end{equation}
where $\vec{v}_i$ is the velocity of the particle before the collision
step, $\Delta \vec{p}/m_i=\vec{v}_i(t+\Delta t)-\vec{v}_i(t)$ and
$\vec{r}_i$ is the particle position referred to the center of the
cell. The expression eqn.~(\ref{eq:stress_tensor}) for the stress
tensor is given, e.g., in Ref.~\cite{winkler_2009}.

\begin{figure} [bht!]
  \epsfig{file=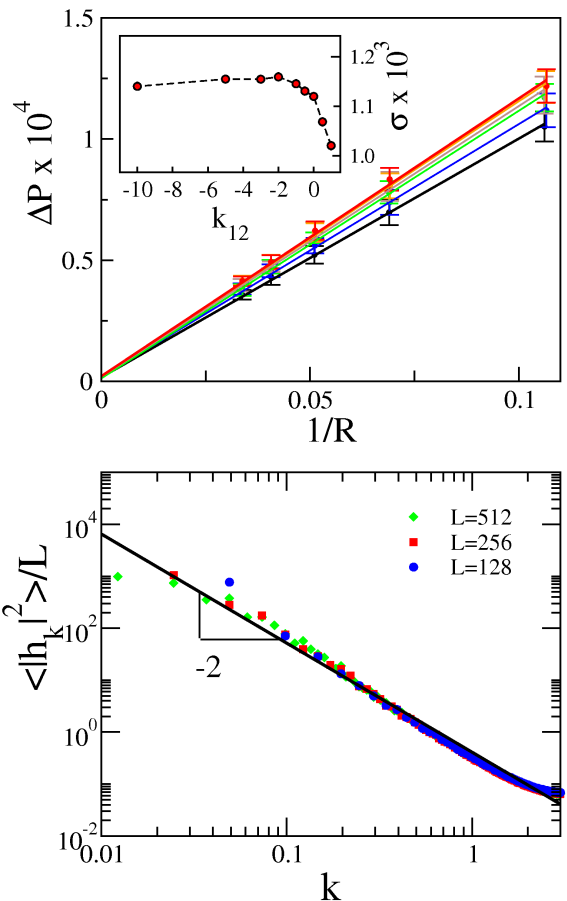,width=0.4\textwidth}
  \caption{Interfacial tension analysis. \textit{Top:} Laplace
      $\Delta P=P_1-P_2$ plotted \textit{vs.} the interfacial
      curvature $1/R$ for droplets of different size. In the inset we
      plot the interfacial tension $\sigma$ estimated from the slope
      of the pressure/curvature plot \textit{vs.} $\kappa_{12}$.
      \textit{Bottom:} Power spectrum of thermally excited capillary
      wave amplitude for a planar interface. The spectra for different
      system sizes collapse according to eqn.~(\ref{eq:equipartition})
      for $\sigma=(0.0023\pm 0.0008)\,m/\Delta
      t$, in agreement with the estimate from the Laplace pressure.
     }
  \label{fig:interfacial_tension}
\end{figure}

Plotting $\Delta P$ against the radius of the averaged droplet contour
in Fig.~\ref{fig:interfacial_tension} shows the expected linear
relation $\Delta P \propto 1/R$ according to Laplace's law. The
proportionality constant is the value of the interfacial tension
$\sigma$. We observe that the interfacial tension becomes non-zero as
soon as $\kappa_{12}\neq \kappa_{11}=\kappa_{22}$.  After a short rise
while decreasing $\kappa_{12}$ , it describes a plateau for values
$\kappa_{12}\lesssim -1$. According to this, in forthcoming
simulations we fix our fluid-fluid interaction $\kappa_{12}=-1$, for
which we have an interfacial tension $\sigma=(0.0011\pm 0.0004)\,m/\Delta t$.

Finally, we check that the thermal fluctuations are consistent with
the estimated value of the interfacial tension analyzing the capillary
waves observed on a planar fluid-fluid interface.  An initially
straight interface will roughen by the motion of thermally activated capillary
waves. In the capillary wave spectrum, each Fourier component $h_k=\sum_{j=0}^{L-1} h_j(t) e^{2\pi ijk/L}$
of the interface displacement contributes according to the equipartition theorem
\begin{equation}
  \langle\lvert h_k \lvert^2 \rangle = \frac{K_BT L}{\sigma k^2}
  \label{eq:equipartition}
\end{equation}
leading to an interface roughness proportional to
$\sqrt{k_BT/\sigma}$~\cite{safran_book,flekkoy_1995}. In Fig.~\ref{fig:interfacial_tension}
we plot the average amplitudes
$\langle\lvert h_k \lvert^2 \rangle$ of the planar capillary waves as
function of the wave number $k$. As expected from the assumption
  of a wave number independent of the interfacial tension $\sigma$ we observe
  that the spectra for different system sizes collapse for the value
  $\sigma=(0.0023\pm 0.0008) m/\Delta t$, which is in good agreement
  with the estimate obtained from the Laplace law for circular droplets
  of different radii. Therefore, the model displays the correct
thermodynamic behavior and interfacial fluctuations.

\section{Flow in porous media}
\label{sec:flow_in_porous_media}

Immiscible displacements in porous media with both capillary and
viscous effects can be characterized by two dimensionless numbers, the
ratio ${\rm M}=\mu_{\rm i}/\mu_{\rm d}$ of the dynamic viscosities
$\mu$ of the \textit{invading} fluid~(i) and the \textit{displaced}
fluid~(d), and the imposed capillary number, i.e., the ratio of
viscous to capillary forces given by ${\rm Ca}=\sigma\,v/\mu_{\rm i}$,
where $\sigma$ is the interfacial tension of the fluid interface and
$v$ is the average front velocity.

When viscous forces dominate (${\rm Ca}>1$), the invading front
exhibits either \textit{stable advance} or \textit{viscous fingering}
(${\rm M}>1$ and ${\rm M}<1$ respectively). However, at low ${\rm Ca}$
the displacement is solely governed by capillary forces~\cite{lenormand_1988,frette_1997,theodoropoulou_2005,berejnov_2007,cottin_2011}.
This propagation mechanism can be captured using concepts from invasion-percolation~\cite{wilkinson_1983,lenormand_1990,cottin_2010,blunt_2004,ferer_2007,tora_hansen_2012}.
In the following, we consider a porous matrix
which is completely filled with the initial fluid phase~(d). By means
of an applied pressure gradient a secondary fluid phase~(i) is pushed
into the matrix displacing phase~(d).

\subsection{System setup}
\label{subsec:system_setup}

The porous matrix is represented by an assembly of non-overlapping
circular disks with fixed position whose preferential wettability by
the invading fluid~(i) can be selected from the three different cases:
\textit{non-wetting, neutral} and \textit{wetting}. Our porous region
is a two-dimensional rectangular channel with length $L$ and width $H$
that extents $0<x<L$ in the direction of the applied pressure
gradient. Periodic boundary conditions are considered into orthogonal
direction.

To generate the pressure gradient that drives the flow we define an
inflow region $x \in [-I,0]$ being free of obstacles where a constant
body force $g$ acts into the x direction on the particles. The
extension of this inflow region is $I\approx0.3L$ into the
direction of the flow. As a result of the body force in the inflow
region a pressure gradient builds up that drives the fluid particles
through the obstacles.

The value of $g$ is constrained to be small enough in order to avoid
compression of the fluids, i.e. a Mach number ${\rm Ma}=v/c<1$, where
$c$ is the speed of sound. In the following we will apply
$g=O(10^{-5})$, which gives us ${\rm Ma}=O(10^{-2})$, Reynolds numbers
${\rm Re}=\rho v \mathcal{L}/\mu=O(10^{-1})$ with $\mathcal{L}$ the
typical distance between the obstacles and mass density $\rho$, and
${\rm Ca}=O(10^{-1})$ for the system sizes and the set of parameters
we consider, as introduced in section~\ref{sec:system_parameters_and_thermal_fluctuations}.

\subsection{Homogeneous wettability}
\label{subsec:homogeneous_wettability}

\begin{figure*}[thb]
  \begin{center}
    \epsfig{file=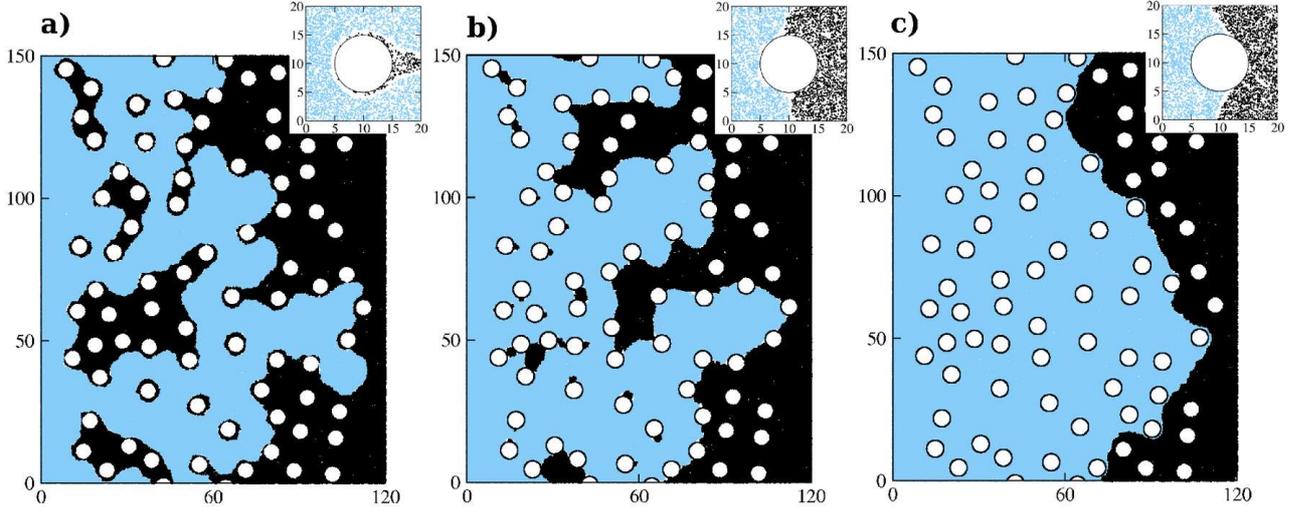,width=0.92\textwidth}
    \caption{Snapshots of the invading front for the three cases of
      homogeneous wettability: a) \textit{non-wetting}, b)
      \textit{neutral wettability} and c) \textit{wetting}. The inset
      of each panel shows a detail of the dynamics around one single
      obstacle. The different advancing mode of the front can be
      appreciated.}
    \label{fig:snapshots_invading_front}
  \end{center}
\end{figure*}

First we analyze the dynamics of the forced displacement in a
substrate composed of monodisperse disks. Previous experiments in quasi-2d
micromodels~\cite{lenormand_1990,cottin_2011}, as well
as numerical approaches~\cite{blunt_2004,ghassemi_2011} clearly
indicate that wetting conditions strongly influence the displacement
of immiscible fluids through porous media in the capillary fingering
regime.

These analyses also evidence that, when viscous forces are not
strictly zero, the flow may be influenced by the pore geometry. For
instance, in the case of a broad throat-size distribution the viscous
pressure drop may become on the order of magnitude of the difference
of Laplace pressure between a small and a large throat, what makes the
invasion-percolation description not longer
valid~\cite{cottin_2010,cottin_2011,hilfer_1998}.

To discern the effect of wettability from effects related to pore
geometry and from dynamic front instabilities (i.e., the
Saffman-Taylor instability) we strictly consider equal dynamic
viscosities ${\rm M}=1$ and porous matrices with low packing density
$\phi=A_o/A$, where $A_o$ is the area occupied by the obstacles an $A$
is the total area of the simulation box. In the following we employ a random configuration of circular
obstacles with radius $r$, generated by means 
of \textit{random sequential addition}~\cite{feder_1980}, for which we impose
$\phi=0.15$ and a minimal separation distance $d=2r$ between the
obstacles.

Figure~\ref{fig:snapshots_invading_front} displays snapshots of the
invading front for the three different cases of wettability, as well
as details of the flow around one single obstacle for each case.
Initially the matrix is completely filled with phase~(d) and then the
invading fluid~(i) is driven from the left by means of the pressure
gradient generated in the inflow region. The applied body force on particles
within this region is $g=1.0\times 10^{-5}$, what generates a
pressure gradient $\Delta P\approx 0.008\, m/\Delta t^2 $.
In these figures we show snapshots of the
invading front close to percolation. They correspond to different
simulation times as the velocity of invasion clearly depends on the
wettability.

\begin{figure}[thb]
  \begin{center}
    \epsfig{file=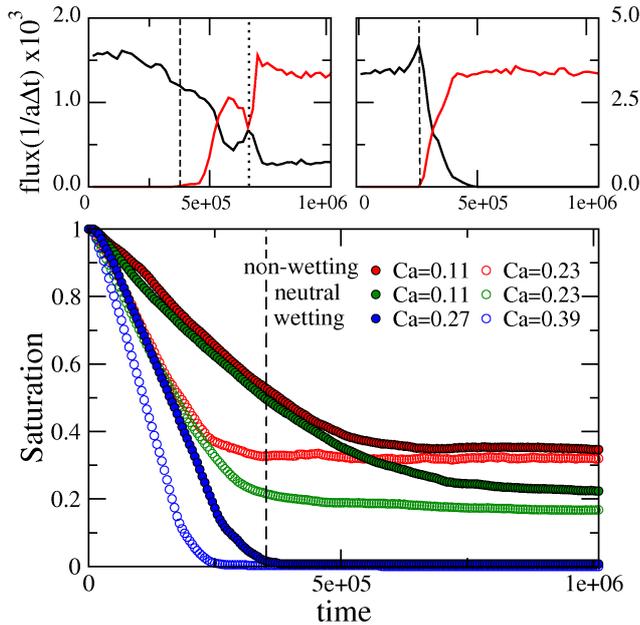,width=0.45\textwidth}
    \caption{Saturation of the displaced phase vs.~time for
      homogeneous wettability: non-wetting \textit{(red)}, neutral
      \textit{(green)} and wetting \textit{(blue)}.  Results are shown
      for low driving $g=1.0\times 10^{-5}$ (full symbols) and higher driving $g=1.5\times 10^{-5}$
      (open symbols). The capillary number for each case is specified in the inset.
      Dashed line is depicted for comparison with Fig.~\ref{fig:fluid-fluid_interface}.
      \textit{Top:} Evolution of the flux of the invading
      \textit{(black)} and displaced fluid phase \textit{(red)} during
      the invasion for the case of non-wetting (left) and wetting
      (right) pore walls for the low driving case. The dashed and dotted lines indicate
      breakthrough and secondary percolation, respectively.}
    \label{fig:homogeneous_saturation_flux}
  \end{center}
\end{figure}

In the non-wetting case we observe that, almost from the very
beginning, the front roughens and fingers are formed. The fluid in
these fingers pinches off leaving behind a number of disconnected
domains of the displaced fluid which are trapped between the disks.
This gives rise to a considerably high residual saturation as we
observe in Fig.~\ref{fig:homogeneous_saturation_flux}, where the evolution of saturation
$S=A/A_0$ is depicted versus time, with the area $A$ of displaced
phase~(d) and total area $A_0$ of the void space. On the other hand,
the formation of new fluid-fluid interface slows down the dynamics, as
we observe when plotting the flux measured at the end of the porous
matrix for both fluid phases. The flux of the invading fluid, instead
of remaining constant in time, as would correspond to a stable front,
clearly decreases in time as the invading fluid describes intricate
paths through the porous structure. These plots also shows that even
after breakthrough, when the flux of the displaced fluid displays a
first peak, some other fingers still develop, and we observe a
secondary percolation of the invading fluid before reaching the steady
state at which the flux becomes constant.

\begin{figure}[tbh]
  \begin{center}
    \epsfig{file=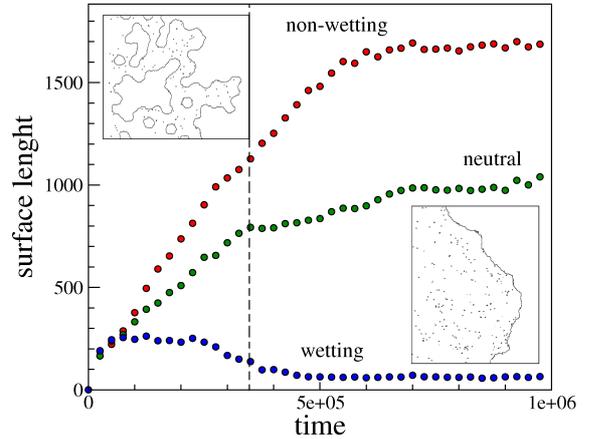,width=0.4\textwidth}
    \caption{Evolution of the fluid-fluid interfacial length in time for the three different homogeneous 
wetting conditions considered: non-wetting, neutral and wetting from top to bottom.
In the insets we show a detail of the fluid-fluid interface for the non-wetting (top) and wetting (bottom) cases,
corresponding to the snapshots presented in Fig~\ref{fig:snapshots_invading_front}.
Dashed line is depicted for comparison with Fig.~\ref{fig:homogeneous_saturation_flux}.}
    \label{fig:fluid-fluid_interface}
  \end{center}
\end{figure}

In the case of neutral wettability the dynamics of the invading front is considerably different.
We observe much smaller trapped clusters of the displaced fluid. A similar behavior has been reported
in Ref~\cite{cottin_2011} for quasi-2d experiments.  This is clearly
manifested by the lower value of residual saturation in
Fig.~\ref{fig:homogeneous_saturation_flux}.  However, in this graph we observe that the velocity
of invasion, that we define as the velocity of an equivalent stable 
front with $v \approx LdS/dt$ for the same imposed ${\rm Ca}$, is very similar for both the
non-wetting and the neutral case until the curves saturate.  As in
both situations the work injected into the system can be invested
either in viscous dissipation or in formation of new fluid-fluid
interfaces, this means that the length of the fluid-fluid interface
should be also comparable in both cases during this time span.  To
figure this out and complement this analysis, we depict in
Fig.~\ref{fig:fluid-fluid_interface} how the length of the fluid-fluid interface evolves
in time. Certainly, we observe that the length growth of the
fluid-fluid interface is very similarly for both cases during this
time span, after which it practically remains constant for the
non-wetting case while is drastically reduced, practically constant,
for the neutral case.

Finally, we consider the case of perfectly wetting obstacles for the
same imposed pressure gradient as in the previous cases. In this case 
the front advance is clearly stable, and the flux of
the invading phase fluctuates around a constant value as well as the length
of the fluid-fluid interface remains constant until breakthrough.
In this case the time until percolation is much shorter, which is readily 
explained by the gain in wetting energy in addition to the pressure gradient.
After percolation, virtually all the initial saturation is taken out as can be seen in Fig.~\ref{fig:homogeneous_saturation_flux}.

Next we analyze how the front advance depends on the applied driving $g$.
In accordance with experimental
results~\cite{cottin_2010} we observe shorter percolation times and
lower values of the residual saturation for higher driving $g=1.5\times 10^{-5}$.
For comparison with the previous results
at lower $\rm Ca$, we include the evolution of $S$ vs. time in Fig.~\ref{fig:homogeneous_saturation_flux}.
It is clear that, in the case of higher Ca, the capillary fingers can overcome more
easily the local Laplace pressure required to invade the pores, that in
this regime is basically given by the pore geometry (at least in the
limit ${\rm M}=1$). This clearly favors the extraction of initial
fluid, as we can appreciate in Fig.~\ref{fig:homogeneous_saturation_flux}, where the
residual values are smaller than the ones obtained for any case of
wettability at smaller ${\rm Ca}$.

\subsection{Heterogeneous wettability}
\label{subsec:heterogeneous_wettability}

In this section we analyze how wettability heterogeneities at pore
scale affect the dynamics of fluid invasion.
Previous works dealing with heterogeneous wetting conditions are present in the Literature,
for instance experiments dealing with a mixture of wetting and non-wetting beads~\cite{salin_1998} or composite
porous medium made of blocks of the same permeability but opposite wettability~\cite{bertin_1998}.
We find also detailed numerical approaches, like simulations of immiscible displacement through a realistic pore structure whose
wettability distribution is taken from microtomographic images of reservoir rocks~\cite{hazlett_1998}.

However, our aim here is to analyze the effect of such heterogeneities using well-characterized spatial distribution
of wettability down to the pore scale. This implies being able to independently assign to each pore a well
defined wettability according to a predefined spatial distribution.

In order to identify the pores in our porous matrices we consider a Delaunay triangulation based on
the centers of the obstacles. According to the construction no
obstacle-center is inside the circumcircle of any other triangular pore.
Once defined the pores, we independently assign wetting conditions to each of them.
This means that our obstacles are divided in sectors according to
the pore definition, and different wettabilities are assigned to
each of them, to build up an heterogeneous pattern at the pore scale.

\begin{figure}[bht]
  \begin{center}
    \epsfig{file=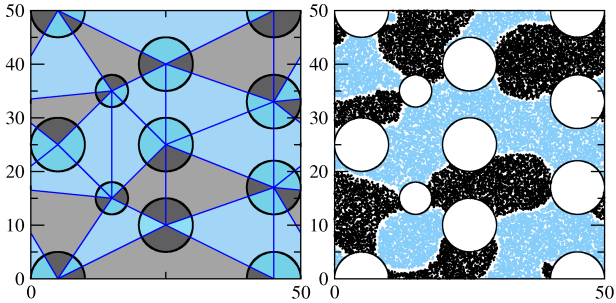,width=0.43\textwidth}
    \caption{Pores defined in terms of Delaunay triangularization
        and wettability definition at pore scale. On the right we show
        the final state of a 1:1 binary mixture of two immiscible fluids
        after relaxation, which matches with the imposed wettability
        pattern defined in terms of wetting and non-wetting pores.}
    \label{fig:delaunay_pores_wettability}
  \end{center}
\end{figure}

Figure \ref{fig:delaunay_pores_wettability} shows the distribution of
two immiscible fluids of an initially random mixture in presence of
a pore-scale wettability pattern after spontaneous phase
separation. It can be observed that each of the two immiscible
fluids is in contact solely to the highly wettable parts of the pore
wall. Deviations from an ideal distribution can be easily explained
by the presence of thermal fluctuations.

Next we consider the same porous geometry than that of 
section~\ref{subsec:homogeneous_wettability} and impose wetting
patterns with well defined spatial correlations, in terms of wetting
and non-wetting domains with typical size $\xi$. To generate these
patterns we start from a randomly distributed wettability for each
pore and introduce correlations by means of a simple stochastic
procedure in which

\begin{itemize}
\item[--] we randomly choose a pair of pores
\item[--] we assign to the second pore the wettability of the first one
  with probability $p(\ell)$ that depends on the distance $\ell$
  between both of them, i.e., the distance between the incenters of the triangles.
\end{itemize}

We choose this function to be the error function $p(\ell)=\frac{1}{2} \left( 1+erf
  \left(\frac{\mu-\ell}{\sigma\sqrt{2}}\right) \right)$, that exhibits
a sharp transition between $1$ and $0$ at a typical distance $\mu$. We fix 
$\sigma=3\mu$.
Due to computational system-size limitations, the number of pores in our system is
not large enough and there is some deviation between the typical distance $\mu$
imposed in the pattern generation and the spatial correlation length $\xi$ 
that the patterns finally exhibit. However, when fitting the spatial correlation 
of the samples to the functional form $C(\ell) \backsim exp(-\ell/\xi)$ we obtain a
correlation length $\xi$ that depends linearly on the $\mu$ parameter as
$\xi=1.2+\mu$ (goodness of fit $C=0.98$). On the other
hand, the ratio the total area occupied by the wetting and non-wetting pores, 
$A_w$ and $A_{nw}$ respectively, is approximately constant for all the patterns
$A_w/A_{n-w}\approx 0.5$.


In the following we will analyze how the variation of the spatial correlation of
our wettability patterns affect the invasion when considering the same arrangement of
obstacles than that of the homogeneous case.

In the case of spatially uncorrelated wettability, i.e. small values of $\mu$,
the front needs to invade many small non-wetting domains in order to percolate.
We show these pore-breakthrough events in Fig.~\ref{fig:pore-breakthroughs} for a given realization
at small domain size with $\mu=4$. For the sake of better
visibility the invaded non-wetting pores are marked in red.

\begin{figure} [H]
  \begin{center}
    \epsfig{file=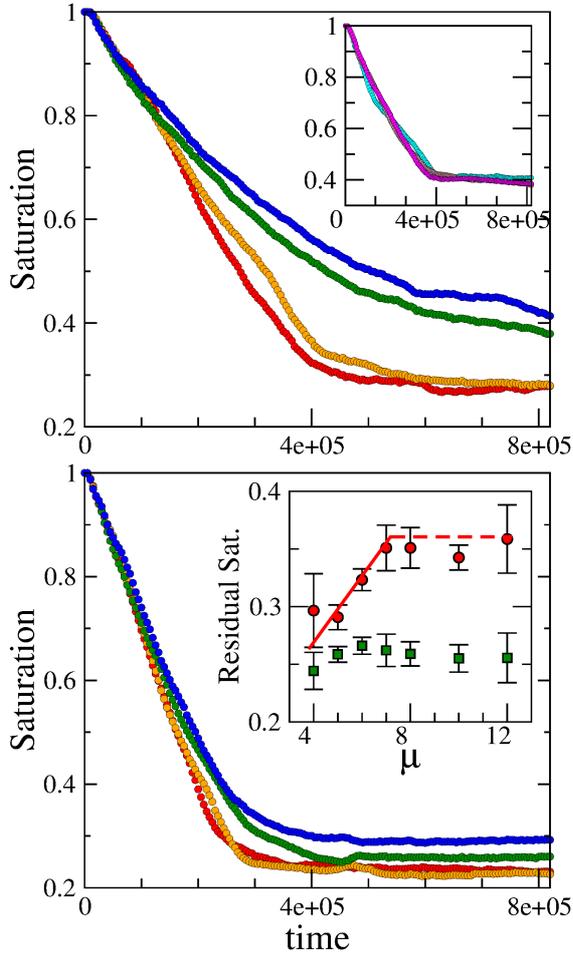,width=0.4\textwidth}
    \caption{Saturation as function of time for heterogeneous
         wettability patterns with different $\mu$ (single realization for each case).
         \textit{Top}: At intermediate correlation lengths, we observe a
         longer time until percolation accompanied by a higher residual
          saturation for increasing $\xi$. The curves correspond to
          parameter $\mu = 4,5,6,7$ (from bottom to top). In the inset the saturation curves
          for larger correlation of the pattern are depicted ($\mu=8,9,12$). In this case
          the invading front rapidly percolates with constant value of residual saturation.
          All the simulations are executed with driving $g=1.0\times 10^{-5}$.
          \textit{Bottom}: Saturation curves for higher imposed $\rm Ca$ ($g=1.5\times 10^{-5}$),         
          for the same parameters values $\mu=4,5,6,7$. The higher applied pressure gradient washes out
          the effect of the wettability patterns, and both the invading velocity and the 
        residual saturation become similar for all the cases. \textit{Inset}: Residual saturation \textit{vs.} $\mu$
        averaged over several equivalent patterns for each value of $\mu$ and independent realizations on each of them.
        Circles and squares correspond to the lower and higher imposed $\rm Ca$, respectively. The higher driving clearly
        diminishes the effect of the wettability pattern. Lines are included to guide the eye: $y\sim 0.03\mu$ (solid) and $y\sim const.$ (dashed)}
    \label{fig:heterogeneous_saturation}
  \end{center}
\end{figure}

\begin{figure*}[tbh!]
\begin{center}
  \epsfig{file=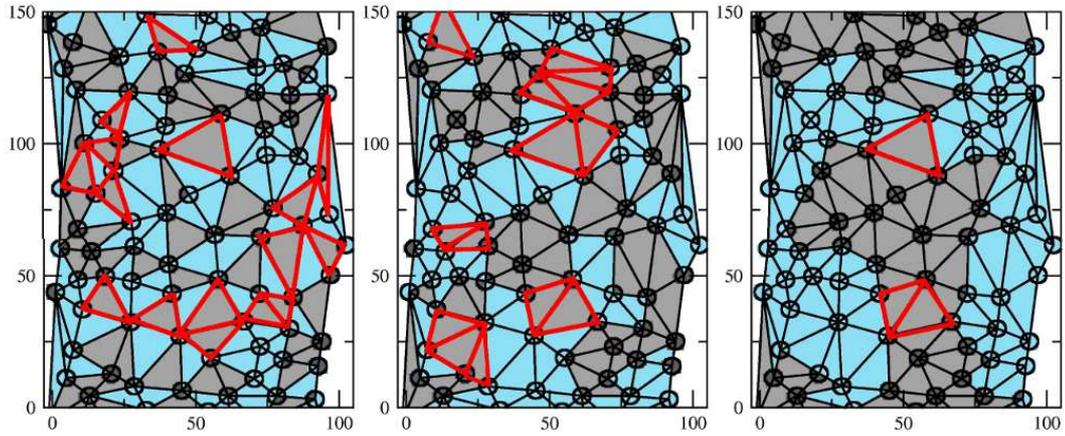,width=0.75\textwidth}
  \caption{Pore invasion for the same single realizations presented in Fig.~\ref{fig:heterogeneous_saturation} with
    spatial correlations of pores wettability corresponding to $\mu=4,6,12$ (from left
    to right). The invaded non-wetting pores are marked in red. The
    invasion process is clearly dominated by the geometrical
    disposition of the wetting patterns.  At low correlation the front
    invades many small non-wetting domains, and the size of these invaded non-wetting domains increases with the correlation length.
    However, for high enough correlations only few single
    bottle-neck non-wetting pores are invaded, and the front rapidly
    percolates.}
  \label{fig:pore-breakthroughs}
\end{center}
\end{figure*}

If we increase the correlations, percolation of the invading fluid
requires filling of larger non-wetting domains.
This slows down the invading process as can be seen
in Fig.~\ref{fig:heterogeneous_saturation}, where we show the evolution of saturation $S$ for 
different values of the $\mu$ parameter. On the other hand, invading these larger
non-wetting domains is more expensive in terms of surface energy, and
therefore many of the non-wetting domains remain unexplored. As we
see from the plots shown in Fig.~\ref{fig:heterogeneous_saturation}, the residual
saturation increases monotonously with the correlation length for this
range of intermediate correlations.

Finally, for even larger correlation lengths we observe that only
few bottle-neck non-wetting pores need to be invaded to achieve
  percolation and the invading fluid rapidly floods the large wetting
domains. Now, the residual saturation is virtually given by the
ratio $A_w/A_{n-w}$ and does not depend any more on
the correlation length. 

We have to mention here that these results are statistically poor due
to the computational limitations to go for larger systems and have better
defined spatial correlations. However, they are still good enough
to show qualitatively that the fluid invasion is sensitive to the spatial 
distribution of wettability at pore scale. In Fig.~\ref{fig:heterogeneous_saturation} we show the dependence of the 
residual saturation after the invasion on the correlation parameter $\mu$, where each value
has been averaged over several equivalent samples and independent realizations on each of them. 
The increase of the residual saturation with the correlations of the wettability patterns 
is clearly appreciable, as well as the subsequent plateau due to the presence of bottle-neck non-wetting
pores.

Finally, we conclude this analysis increasing the imposed ${\rm Ca}$, as we
did in the case of homogeneous wettability. Now the higher pressure gradient washes out
the effect of the wettability patterns on the flow advance. We appreciate 
in Fig.~\ref{fig:heterogeneous_saturation} that the saturation curves for different $\mu$ are similar
and the difference in residual saturation diminishes, in comparison with the lower driving case.

\section{Conclusions}

We have adapted the MPC multicolor algorithm to control the wetting
conditions at the solid boundaries in order to simulate immiscible
displacement trough porous media. This is  an alternative
particle-based simulation approach that provides microscopical
details of the dynamics at the fluid boundaries and directly incorporates thermal
fluctuations.

As a first application of this method, we analyze the dependence of the
multiphase dynamics on the spatial correlations of wettability,
analyzing the evolution of the initial saturation and particle flux
for different wetting conditions.

In the case of homogeneous wettability we observe that the residual
saturation decreases with the affinity of the invading fluid to the
solid phase, as well as with the imposed ${\rm Ca}$ for every wetting
condition.

The role of heterogeneous wettability is analyzed by means of
different prescribed spatial correlations at the pore scale.
We observe that the dynamics is influenced by such spatial correlations.
While in the range of intermediate correlations the residual saturation increases with $\xi$
whereas the invading process becomes slower, for large enough
correlations (of the order of the system size) the presence of some
bottle-neck pores controls the dynamics, and the invading fluid rapidly
percolates, while the residual saturation remains constant given by
the ratio of wetting and non-wetting pore area.

These qualitative results manifest the interest of a further
exploration of the influence of the wettability distribution at the scale of the pore.
In this work, the statistics of our heterogeneous patterns is limited by the system size, that
is in turn limited by the computational costs of the MPC algorithm implemented in sequential mode.
However,this algorithm provides a good level of parallelism. In the free streaming step, the particles
are propagated according to their velocity without interacting with each other. 
On the other hand, the multi-particle collision step  is performed cellwise.
This means that the necessary calculations are completely independent from cell to cell, so that they 
can be also executed in parallel. This will be certainly exploited in further works in order
to carry out more accurate simulations to analyze microscopic flow
mechanisms that are observed at pore scale. On the other hand, this MPC
multicolor model has the advantage of being able to deal with an arbitrary
number of phases, what is a interesting feature to simulate more realistic systems.

We thank Yasuhiro Inoue and Anikka Schiller for helpful discussions as well as 
Stephan Herminghaus for his comments and support. Funding from BP Exploration Operating Company Ltd. within the ExploRe research program
is gratefully acknowledged.

\bibliographystyle{unsrt}  
\bibliography{references}  

\end{document}